\documentclass[twocolumn,showpacs,preprintnumbers,amsmath,amssymb, showkeys]{revtex4}

\usepackage{graphicx}
\usepackage{dcolumn}
\usepackage{bm}
\usepackage{epsf}

\begin{document}


\title{Non-stationary effects in the coupled quantum dots influenced by the electron-phonon interaction}

\author{V.\,N.\,Mantsevich}
 \altaffiliation{vmantsev@spmlab.phys.msu.ru}
\author{N.\,S.\,Maslova}%
 \email{spm@spmlab.phys.msu.ru}
\author{P.\,I.\,Arseyev}
 \altaffiliation{ars@lpi.ru}
\affiliation{Moscow State University, Department of  Physics, 119991
Moscow, Russia\\~\\ P.N. Lebedev Physical institute of RAS, 119991,
Moscow, Russia}

\date{\today }
6 pages, 2 figures
\begin{abstract}
We analyzed time evolution of the localized charge in the system of
two interacting single level quantum dots (QDs) coupled with the
continuous spectrum states in the presence of electron-phonon
interaction.

We demonstrated that electron-phonon interaction leads to the
increasing of localized charge relaxation rate. We also found that
several time scales with different relaxation rates appear in the
system in the case of non-resonant tunneling between the dots. We
revealed the formation of oscillations in the filling numbers time
evolution caused by the emission and adsorption processes of
phonons.
\end{abstract}

\maketitle

\section{Introduction}

Recent progress in the engineering and fabrication of well-defined
artificial systems -  quantum dots (QDs), leads to the possibility
of ultra small electronic devices formation with a relatively high
control of system parameters (size, shape and energy spectrum)
\cite{Tan}. In addition to the QDs potential industrial
applications, these nanoscale objects provide an ideal test ground
for the study of basic physics including, many-body interaction
effects, electron transport and time-dependent effects. Moreover,
QDs integration in a little quantum circuits deals with careful
analysis of non-equilibrium charge distribution, relaxation
processes and non-stationary effects. These processes influence
strongly on the electron transport through the system of QDs
\cite{Rasmussen},\cite{Loss},\cite{Arseyev},\cite{Arseyev_1},\cite{Arseyev_2},\cite{Contreras}.
Electron transport in such systems is governed not only by the
Coulomb interaction between localized electrons
\cite{Arseyev_1},\cite{Arseyev_2},\cite{Contreras} but also by the
electron-phonon interaction \cite{Comas},\cite{Keil},\cite{Zhu}.

Most of the theoretical works devoted to the problem of electron
transport through the coupled QDs in the presence of electron-phonon
interaction deal with the tunneling current and current-current
correlations (shot noise) investigations \cite{Zhu},\cite{Dong}.
Only a few attempts have been made to analyze phonon assisted
localized charge relaxation
\cite{Wu},\cite{Stavrou},\cite{Climente}. It was found theoretically
that lateral confinement influence on the single-electron relaxation
rates in parabolic QDs \cite{Bockelmann}. Author considered the
deformation potential coupling between electrons and longitudinal
acoustic phonons, while neglecting piezoelectric coupling on the
grounds of it's weaker contribution in two dimensional structures.
On the other hand further theoretical and experimental works
suggested that electron-acoustic phonon scattering due to the
piezoelectric field interaction is relevant for momentum and spin
relaxation processes \cite{Fujisawa},\cite{Cheng} and may even
provide charge decoherence in laterally coupled QDs
\cite{Wu},\cite{Stavrou}. In \cite{Climente} authors analyzed phonon
induced single electron relaxation rates in the models of weakly
confined single and vertically coupled QDs taking into account both
mentioned above mechanisms. The regimes where each coupling
mechanism prevails were found.

In this paper we use the Keldysh diagram technique \cite{Keldysh} to
analyze charge relaxation in a double QDs due to the coupling with
the continuous spectrum states in the presence of electron-phonon
interaction. Tunneling to the reservoir is possible only from one of
the dots. We have found that electron-phonon interaction results in
the increasing of localized charge relaxation rate and also leads to
the formation of well resolved oscillations.

\section{The suggested model}
 In the present paper we consider a system of coupled QDs with the
single particle levels $\varepsilon_1$ and $\varepsilon_2$ connected
with the continuous spectrum states. At the initial time two
electrons with opposite spins are localized in the first QD on the
energy level $\varepsilon_1$ ($n_{1\sigma}(0)=n_{0}=1$). The second
QD with the energy level $\varepsilon_2$ is connected with the
reservoir ($\varepsilon_p$). Relaxation of the localized charge is
governed by the Hamiltonian:

\begin{eqnarray}
\Hat{H}=\Hat{H}_{D}+\Hat{H}_{tun}+\Hat{H}_{res}.
\end{eqnarray}

The Hamiltonian $\Hat{H_{D}}$ of interacting QDs

\begin{eqnarray}
\Hat{H}_{D}&=&\sum_{i=1,2\sigma}\varepsilon_{i}c^{+}_{i\sigma}c_{i\sigma}+\sum_{\sigma}T(c_{1\sigma}^{+}c_{2\sigma}+c_{1\sigma}c_{2\sigma}^{+})+\nonumber\\&+&\omega_{0}b^{+}b+g(c_{1\sigma}^{+}c_{2\sigma}+c_{1\sigma}c_{2\sigma}^{+})(b^{+}+b)
\end{eqnarray}

contains the spin-degenerate levels $\varepsilon_i$ (indexes $i=1$
and $i=2$ correspond to the first and to the second QD) and
electron-phonon interaction. The creation/annihilation of an
electron with spin $\sigma=\pm1$ within the dot is denoted by
$c^{+}_{i\sigma}/c_{i\sigma}$ and $n_{\sigma}$ is the corresponding
filling number operator. Operators $b^{+}/b$ describe the
creation/annihilation of the phonons. $\omega_0$ - is the optical
phonon frequency and $g$ - is the electron-phonon coupling constant.
The interaction between the dots is described by the tunneling
transfer amplitude $T$ which is considered to be independent of
momentum and spin.

The coupling between the second dot and the continuous spectrum
states is described by the Hamiltonian:

\begin{eqnarray}
\Hat{H}_{tun}=\sum_{p\sigma}t(c_{p\sigma}^{+}c_{2\sigma}+c_{p\sigma}c_{2\sigma}^{+}),
\end{eqnarray}

where $t$ is the tunneling amplitude, which we assume to be
independent on momentum and spin. By considering a constant density
of states in the reservoir $\nu_0$, the tunnel rate $\gamma$ is
defined as $\gamma=\pi\nu_0t^{2}$.

The continuous spectrum states are modeled by the Hamiltonian:

\begin{eqnarray}
\Hat{H}_{res}=\sum_{p\sigma}\varepsilon_{p}c^{+}_{p\sigma}c_{p\sigma},
\end{eqnarray}

where $c^{+}_{p\sigma}/c_{p\sigma}$ creates/annihilates an electron
with spin $\sigma$ and momentum $p$ in the lead. We shall use
Keldysh diagram technique to describe charge density relaxation
processes in the considered system. Time evolution of the electron
density in the QD is determined by the Keldysh Green function
$G_{11}^{<}$ which is connected with the localized state filling
numbers in the following way:

\begin{eqnarray}
G_{11}^{<}(t,t)=in_{1}(t)
\end{eqnarray}

Integro-differential equations for Green function
$G_{11}^{<T}(t,t^{'})$ without electron-phonon interaction has the
form:

\begin{eqnarray}
G_{11}^{<T}(t,t^{'})&=&G_{11}^{0<}+G_{11}^{0R}T^{2}G_{22}^{0R}G_{11}^{<}+\nonumber\\&+&G_{11}^{0R}T^{2}G_{22}^{0<}G_{11}^{AT}+G_{11}^{0<}T^{2}G_{22}^{0A}G_{11}^{AT}\nonumber\\
\label{G_0}
\end{eqnarray}

In the case when initial charge is localized in the first QD and the
second dot is empty, the third term in the eq.(\ref{G_0}) can be
neglected. Retarded Green's function
$G_{11}^{AT}(t^{'},t)=[G_{11}^{RT}(t,t^{'})]^{*}$ yields density of
states in the first QD and can be found exactly from the integral
equation:

\begin{eqnarray}
G_{11}^{RT}&=&G_{11}^{0R}+G_{11}^{0R}T^{2}G_{22}^{0R}G_{11}^{R}
\label{integral_equation}
\end{eqnarray}

where Green's functions $G_{11}^{0R}(t-t^{'})$ and
$G_{22}^{0R}(t-t^{'})$ in the absence of coupling between the dots
are determined by the expressions:

\begin{eqnarray}
G_{11}^{0R}(t-t^{'})&=&-i\Theta(t-t^{'})e^{-i\varepsilon_1(t-t^{'})}\nonumber\\
G_{22}^{0R}(t-t^{'})&=&-i\Theta(t-t^{'})e^{-i\varepsilon_2(t-t^{'})-\gamma(t-t^{'})}
\end{eqnarray}

The eigenfrequencies $E_{1,2}$ of equation (\ref{integral_equation})
are determined in the following way

\begin{eqnarray}
(E-\varepsilon_1)(E-\varepsilon_2+i\gamma)-T^{2}=0\nonumber\\
E_{1,2}=\frac{1}{2}(\varepsilon_1+\varepsilon_2-i\gamma)\pm\frac{1}{2}\sqrt{(\varepsilon_1-\varepsilon_2+i\gamma)^{2}+4T^{2}}
\end{eqnarray}

Finally retarded Green's function can be written as:

\begin{eqnarray}
G_{11}^{RT}(t,t^{'})=-i\Theta(t-t^{'})(\frac{E_1-\varepsilon_2+i\gamma}{E_1-E_2}e^{-E_{1}(t-t^{'})}-\nonumber\\
-\frac{E_2-\varepsilon_2+i\gamma}{E_1-E_2}e^{-E_{2}(t-t^{'})})\label{G}
\end{eqnarray}

and interaction with the continuous spectrum states is included in
the Green's function $G_{22}^{0R}(t-t^{'})$. Electron-phonon
interaction results in the appearance of corrections to the Green's
function $G_{11}^{RT}$ in the equations (\ref{G_0}) and
(\ref{integral_equation}). Consequently the equation for Green
function has the following form:

\begin{eqnarray}
G_{11}^{R}(t,t^{'})&=&G_{11}^{0R}+G_{11}^{0R}T^{2}G_{22}^{0R}G_{11}^{R}+G_{11}^{0R}\Sigma_{11}^{R}G_{11}^{R}+\nonumber\\&+&G_{11}^{0R}\Sigma_{12}^{R}G_{21}^{R}
+G_{12}^{0R}\Sigma_{21}^{R}G_{11}^{R}+G_{12}^{0R}\Sigma_{22}^{R}G_{21}^{R}
\label{G_{11}^{R}}\nonumber\\
\end{eqnarray}

where self-energies $\Sigma_{11}^{R}(t,t^{'})$,
$\Sigma_{12}^{R}(t,t^{'})$, $\Sigma_{21}^{R}(t,t^{'})$ and
$\Sigma_{22}^{R}(t,t^{'})$ can be written as:

\begin{eqnarray}
\Sigma_{11}^{R}(t,t^{'})&=&ig^{2}[D^{>}G_{22}^{AT}+D^{R}G_{22}^{<T}]\nonumber\\
\Sigma_{12}^{R}(t,t^{'})&=&ig^{2}[D^{>}G_{21}^{AT}+D^{R})G_{21}^{<T}]\nonumber\\
\Sigma_{21}^{R}(t,t^{'})&=&ig^{2}[D^{>}G_{12}^{AT}+D^{R}G_{12}^{<T}]\nonumber\\
\Sigma_{22}^{R}(t,t^{'})&=&ig^{2}[D^{>}G_{11}^{AT}+D^{R}G_{11}^{<T}]
\label{qwe}
\end{eqnarray}

In the eq.(\ref{qwe}) the following ratio is considered
$G_{22}^{AT}(t^{'},t)=[G_{22}^{RT}(t,t^{'})]^{*}$ and expression for
Green's function $G_{22}^{RT}(t,t^{'})$ analogous to the equation
(\ref{G}) has the following form:

\begin{eqnarray}
G_{22}^{RT}(t,t^{'})=-i\Theta(t-t^{'})(\frac{E_2-\varepsilon_1}{E_1-E_2}e^{-E_{1}(t-t^{'})}-\nonumber\\
-\frac{E_1-\varepsilon_1}{E_1-E_2}e^{-E_{2}(t-t^{'})})
\end{eqnarray}

The last three terms in eq.(\ref{G_{11}^{R}}) correspond to the next
order perturbation theory in the parameter
$\frac{T^{2}}{\gamma^{2}}$. Consequently, localized charge
relaxation in the presence of electron-phonon interaction is mostly
governed by the term $G_{11}^{0R}\Sigma_{11}^{R}G_{11}^{R}$. Acting
with operator $G_{11}^{0R-1}$ and considering the terms with the
accuracy $\frac{T^{2}}{\gamma^{2}}$ one can re-write the equation
(\ref{G_{11}^{R}}) in the following way:

\begin{eqnarray}
(G_{11}^{0R-1}-T^{2}G_{22}^{0R}-\Sigma_{11}^{R})G_{11}^{R}(t,t^{'})&=&\delta(t-t^{'})\label{G_{1}^{R}}\end{eqnarray}

The eigenvalues of equation (\ref{G_{1}^{R}}) can be found from
characteristic equation written in the following form:

\begin{eqnarray}
[G_{11}^{0R-1}(\omega)G_{22}^{0R-1}(\omega)-T^{2}]\cdot\nonumber\\
\cdot[G_{11}^{0R-1}(\omega-\omega_{0})G_{22}^{0R-1}(\omega-\omega_{0})-T^{2}]-\nonumber\\-g^{2}\cdot(2N_{0\omega}+1)\cdot
G_{22}^{0R-1}(\omega)G_{11}^{0R-1}(\omega-\omega_{0})=0\nonumber\\
\label{eigenvalues}
\end{eqnarray}

where functions $G_{ii}^{0R-1}$ can be determined as

\begin{eqnarray}
G_{ii}^{0R-1}&=&i\frac{\partial}{\partial t}-\varepsilon_i\nonumber\\
\end{eqnarray}

Consequently retarded Green's function $G_{11}^{R}$ can be written
in the following form:

\begin{eqnarray}
G_{11}^{R}(t,t^{'})=\sum_{i}-i\Theta(t-t^{'})A_{i}e^{-iE_{i}(t-t^{'})}
\label{green_function}\end{eqnarray}

where $E_{i}$ - are eigenvalues of equation (\ref{eigenvalues}).
Coefficients $A_{i}$ can be found from the system of linear
equations obtained in the first order perturbation theory in the
parameter $g^{2}$:

\begin{eqnarray}
\sum_{i=1}^{4}A_i&=&1\nonumber\\
-\sum_{i=1}^{4}A_i\cdot\sum_{j\neq i}E_j&=&-(E_{3}^{0}+E_{4}^{0}+\varepsilon_2-i\gamma)\nonumber\\
\sum_{i=1}^{4}A_i\cdot\sum_{k\neq l\neq i}E_k\cdot E_l&=&E_{3}^{0}E_{4}^{0}+(\varepsilon_2-i\gamma)(E_{3}^{0}+E_{4}^{0})\nonumber\\
\sum_{i=1}^{4}A_i\cdot \prod^{j\neq i}
E_j&=&-(\varepsilon_2-i\gamma)E_{3}^{0}E_{4}^{0}
\end{eqnarray}

$E_{i}^{0}$ - are the eigenvalues of equation (\ref{eigenvalues})
with electron-phonon coupling constant $g=0$.

\begin{eqnarray}
E_{1,2}^{0}&=&E_{1,2}\nonumber\\
E_{3,4}^{0}&=&\omega_0+E_{1,2}
\end{eqnarray}

Equation for Keldysh Green function $G_{11}^{<}(t,t^{'})$, which
determines localized charge time evolution $n_{1}(t)$ than has the
form:

\begin{eqnarray}
G_{11}^{<}(t,t^{'})=G_{11}^{0<}+G_{11}^{0R}T^{2}G_{22}^{0R}G_{11}^{<}+\nonumber\\+G_{11}^{0R}T^{2}G_{22}^{0<}G_{11}^{A}+G_{11}^{0<}T^{2}G_{22}^{0A}G_{11}^{A}+\nonumber\\+G_{11}^{0R}\Sigma_{11}^{<}G_{11}^{A}+G_{11}^{0R}\Sigma_{11}^{R}G_{11}^{<}+
G_{11}^{0<}\Sigma_{11}^{A}G_{11}^{A}\label{eq}
\end{eqnarray}

where self-energy $\Sigma_{11}^{<}(t,t^{'})$ can be written as:

\begin{eqnarray}
\Sigma_{11}^{<}(t,t^{'})&=&ig^{2} D^{<}(t,t^{'})G_{22}^{<}(t,t^{'})
\end{eqnarray}

with phonon function $D^{<}(t_1,t_2)$:

\begin{eqnarray}
D^{<}(t_1,t_2)=-i(N_{\omega0}+1)e^{-i\omega_{0}(t_1-t_2)}-iN_{-\omega0}e^{i\omega_{0}(t_1-t_2)}\nonumber\\
\end{eqnarray}

acting with operator $G_{11}^{0R-1}$ equation (\ref{eq}) can be
re-written in the following form:

\begin{eqnarray}
G_{11}^{0R-1}G_{11}^{<}(t,t^{'})=(i\frac{\partial}{\partial
t}-\varepsilon_1)G_{11}^{<}(t,t^{'})=\nonumber\\= T^{2}
\int_{0}^{\infty}dt_{1}\cdot
G_{22}^{0R}(t,t_{1})G_{11}^{<}(t_{1},t^{'}) +\nonumber\\+\int
dt_{1}[\Sigma_{11}^{<}(t,t_{1})G_{11}^{A}(t_{1},t^{'})+\Sigma_{11}^{R}(t,t_{1})G_{11}^{<}(t_{1},t^{'})]
\end{eqnarray}

Green's function $G_{11}^{<}(t,t^{'})=in_{1}(t)$ is determined by
the sum of homogeneous and inhomogeneous solutions:

\begin{eqnarray}
n_1(t)=n_{1}^{h}(t)+\widetilde{n}_{1}(t)=n_{1}^{h}(t)+\nonumber\\+\int_{0}^{t}G_{11}^{R}(t,t_{1})\Sigma_{11}^{<}(t,t_2)G_{11}^{A}(t_2,t)dt_{1}dt_{2}
\end{eqnarray}

Homogeneous solution of the equation can be written in the following
way:

\begin{eqnarray}
n_{1}^{h}(t)=n_{1}^{0}\cdot
\sum_{ij}A^{}_{i}A_{j}^{*}e^{-i(E_{i}-E_{j}^{*})t}
\end{eqnarray}

where coefficients $A_{i}^{}$ correspond to the Green's function
$G_{11}^{R}$, which is determined by the equation
(\ref{green_function}). Function $G_{22}^{<T}(t_1,t_2)$ can be
written as:

\begin{eqnarray}
G_{22}^{<T}(t_1,t_2)=\sum_{i^{'}j^{'}=1,2}a_{i^{'}j^{'}}e^{-iE_{i^{'}}^{0}t_{1}}\cdot
e^{iE_{j^{'}}^{*0}t_{2}}
\end{eqnarray}

and coefficients $a_{i^{'}j^{'}}$ have the following form:

\begin{eqnarray}
a_{11}&=&a_{22}=\frac{iT^{2}}{|E_{2}^{0}-E_{1}^{0}|^{2}}\nonumber\\
a_{12}&=&a_{21}^{*}=-a_{11}
\end{eqnarray}

Consequently one can find the inhomogeneous solution of the equation

\begin{eqnarray}
\widetilde{n}_{1}(t)= g^{2}\sum_{iji^{'}j^{'}=1}^{4}A_{i}^{}\cdot
A_{j}^{*}\cdot a_{i^{'}j^{'}}\cdot\nonumber\\ \frac{-1}{i(E_{j}^{*}-E_{j^{'}}^{0*}-\omega_{0})}\cdot\frac{1}{i(E_{i}^{}-E_{i^{'}}^{0}-\omega_{0})}\cdot\nonumber\\
\cdot[e^{-i(E_{i^{'}}^{0}+\omega_0)t}-e^{-iE_{i}^{}t}][e^{i(E_{j^{'}}^{0*}+\omega_0)t}-e^{iE_{j}^{*}t}]
\label{sys}\end{eqnarray}.

Considering only the leading terms in parameters
$\frac{g^{2}}{\omega_{0}^{2}}$, $\frac{T^{2}}{\gamma^{2}}$ in
eq.(\ref{sys}), the following expression for the inhomogeneous
solution can be obtained:

\begin{eqnarray}
\widetilde{n}_{1}(t)=\frac{g^{2}}{\omega_{0}^{2}}\cdot\frac{T^{2}}{\gamma^{2}}\sum_{i^{'}j^{'}=1}^{2}[e^{-i(E_{i^{'}}^{0}+\omega_0)t}-e^{-iE_{1}^{}t}]\cdot\nonumber\\
\cdot[e^{i(E_{j^{'}}^{0*}+\omega_0)t}-e^{iE_{1}^{*}t}]
\end{eqnarray}

\section{Results and discussion}

The behavior of filling numbers time evolution depends on the
parameters of the system: energy levels detuning, the relation
between tunneling rates and electron-phonon coupling constant, the
value of optical phonon frequency. The general feature of all
dependencies is the increasing of localized charge relaxation rate
caused by the electron-phonon interaction.

\begin{figure}[h]
\centering
\includegraphics[width=70mm]{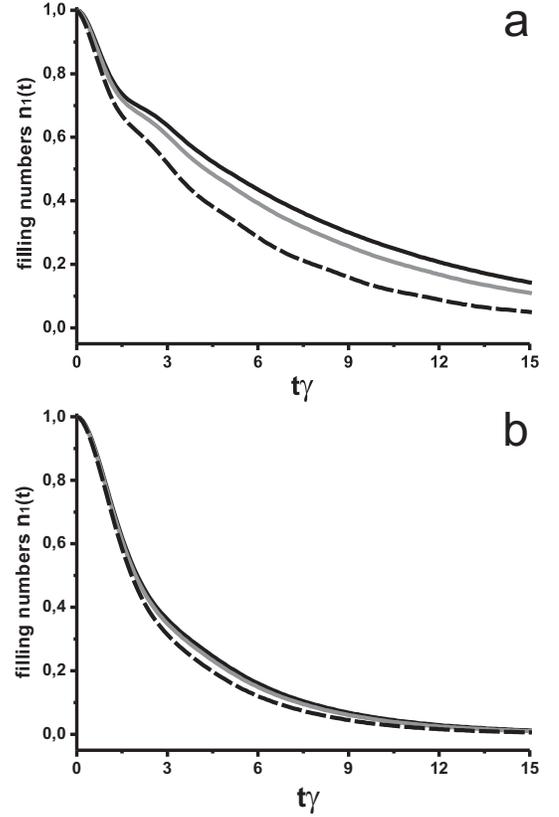}%
\caption{Fig.1 Filling numbers time evolution in the presence of
electron-phonon interaction in the case of positive initial detuning
$\Delta\varepsilon$. Black line corresponds to the case when $g=0$,
grey line describes the situation when $g=0.1$ and black-dashed line
- $g=0,2$. a) $\Delta\varepsilon=2,0$, $\omega_0=2,0$;
b)$\Delta\varepsilon=1,0$, $\omega_0=1,0$. For all the figures
values of parameters $T=0,6$, $\gamma=1,0$ are the same.}
\label{Fig.1}
\end{figure}

We start by discussing the filling numbers time evolution in the
case of the positive initial detuning between energy levels in the
coupled QDs ($\Delta\varepsilon=\varepsilon_1-\varepsilon_2>0$).
Obtained calculation results are presented on the Fig.\ref{Fig.1}.
It is clearly evident that electron-phonon interaction leads to the
increasing of localized charge relaxation rate. The growth of the
electron-phonon coupling constant $g$ for a given set of system
parameters results in the increasing of filling numbers relaxation
rate. With the increasing of the initial detuning the influence of
electron-phonon interaction on the charge time evolution is clearly
pronounced (see Fig.\ref{Fig.1}a).

\begin{figure}[h]
\centering
\includegraphics[width=70mm]{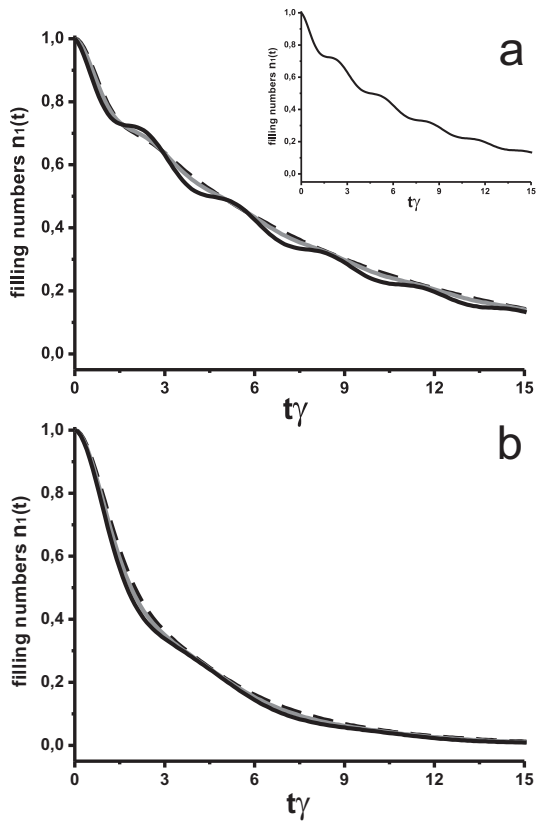}%
\caption{Fig.2 Filling numbers time evolution in the presence of
electron-phonon interaction in the case of negative initial detuning
$\Delta\varepsilon$. Black-dashed line corresponds to the case when
$g=0$, grey line describes the situation when $g=0.1$ and black line
- $g=0,2$. a) $\Delta\varepsilon=-2,0$, $\omega_0=2,0$;
b)$\Delta\varepsilon=-1,0$, $\omega_0=1,0$. For all the figures
values of parameters $T=0,6$, $\gamma=1,0$ are the same. The inset
demonstrates localized charge relaxation in the case when $g=0,2$.}
\label{Fig.2}
\end{figure}

A critical value of the detuning exists in the system under
investigation for a given set of parameters which corresponds to the
relaxation regime changing. For the smaller values of the detuning
charge relaxation takes place with the only one relaxation rate both
in the presence (see grey and black-dashed lines on the
Fig.\ref{Fig.1}b) and in the absence of electron-phonon interaction
(see black line on the Fig.\ref{Fig.1}b). The typical time scale
which determines the localized charge relaxation is close to the
value $\gamma_{res}=2\frac{T^2}{\gamma}$. For the larger values of
detuning than the critical one, localized charge time evolution
reveals two typical time intervals with different values of the
relaxation rates (see Fig.\ref{Fig.1}a). The first time interval
relaxation rate exceeds the relaxation rate of the second time
interval both in the presence and in the absence of electron-phonon
coupling. Without electron-phonon interaction the first time
interval reveals charge relaxation with the typical rate
$\gamma_{res}$. The second time interval demonstrates charge time
evolution with relaxation rate close to
$\gamma_{nonres}=\gamma_{res}\frac{\gamma^{2}}{\Delta\varepsilon^{2}}$.
When electron phonon-coupling is involved, the filling numbers time
evolution in the first time interval has the value
$\gamma\sim2\gamma_{res}$ and in the second time interval  -
$\gamma\sim2\gamma_{nonres}$. Consequently, electron-phonon
interaction results in the two times increasing of localized charge
relaxation rate.

Let us now focus on the charge relaxation processes in the case of
negative initial detuning between energy levels in the QDs
($\varepsilon_1<\varepsilon_2$) (see Fig.\ref{Fig.2}). One can
clearly see that in the case of negative detuning electron-phonon
interaction also results in the increasing of localized charge
relaxation rate, but this effect is less pronounced. Charge time
evolution changes slightly in comparison with the situation when
positive detuning occurs. In the case of negative detuning always
exist several time intervals with different values of the relaxation
rates (see Fig.\ref{Fig.2}). For the small value of initial detuning
(see Fig.\ref{Fig.2}b) relaxation rates on the both time intervals
are very close to each other and to the value $\gamma_{res}$ both in
the presence and in the absence of electron-phonon interaction. For
the larger value of initial detuning in the case when
electron-phonon coupling is absent  the first time interval reveals
charge relaxation with the typical rate $\gamma_{res}$ and the
second time interval corresponds to the relaxation rate
$\gamma_{nonres}$. When electron-phonon coupling is considered,
relaxation rates increase slightly and they are continue being very
close o the values $\gamma_{res}$ and $\gamma_{nonres}$ for the
first and second time intervals correspondingly.

The most interesting effect in this energy levels configuration is
the formation of oscillations in the filling numbers time evolution,
caused by the emission and adsorption of phonons by the energy
levels when detuning is close to the phonon frequency. Oscillations
are well pronounced for the large value of initial detuning (see
Fig.\ref{Fig.2}a and the inset). For a given set of system
parameters the oscillations amplitude is determined by the value of
the electron-phonon coupling constant $g$ (see Fig.\ref{Fig.2}). The
presence of oscillations may even lead to the decreasing of filling
numbers relaxation rate in comparison with the case when
electron-phonon interaction is absent (see black-dashed and black
lines on the Fig.\ref{Fig.2}a) for a given set of system parameters
in the particular time intervals.

\section{Conclusion}

We investigated filling numbers time evolution in the system of two
interacting QDs weakly coupled to the reservoir in the presence of
electron-phonon interaction. It was shown that electron-phonon
interaction results in the increasing of localized charge relaxation
rate. The value of extantion is determined by the system parameters
such as energy levels detuning, optical phonon frequency and the
ratio between electron-phonon coupling constant and tunneling
transfer amplitudes.

We revealed that in the case of positive initial detuning between
energy levels in the dots the influence of electron-phonon
interaction is mostly pronounced and the relaxation rate increases
with the growth of the initial detuning value.

We found that when negative initial detuning is considered, the
influence of electron-phonon interaction leads to the formation of
oscillations in the filling numbers time evolution. These
oscillations are the result of phonons emission and adsorption
between the energy levels.

This work was  supported by RFBR grants and by the National Grants
of Ministry of science and education.


\pagebreak

\end{document}